# Electrodynamical Forbiddance of the Strong Quadrupole Light-Molecule Interaction and Its Experimental Manifestation in Fullerene $C_{60}$

V.P. Chelibanov[a], A.M. Polubotko[b]

[a] The ITMO University, 197101, Saint Petersburg, Russia, E-mail: Chelibanov@gmail.com
[b] A.F. Ioffe Physico-Technical Institute, Russian Academy of Sciences, 194021 Saint Petersburg, Russia E-mail: alex.marina@mail.ioffe.ru

**Abstract**

It is demonstrated that the forbidden lines, which must be present in the SERS, TERS and SEIRA spectra of molecules with sufficiently high symmetry, associated with a strong quadrupole light-molecule interaction, are absent in the fullerene $C_{60}$. This result is an experimental manifestation of an electrodynamical forbiddance of the strong quadrupole light-molecule interaction, which must be not only in molecules with cubic symmetry groups, but in the fullerene $C_{60}$ also.

1. **Introduction**

The theory of SERS, SEHRS and SEIRA is of a great interest since these processes are widely used in science and technology for a sensitive spectral analysis. For a long time it was considered that the reason of these phenomena are plasmons and so-called chemical enhancement. However as it was demonstrated in [1-3] all these phenomena can be explained on the base of so-called strong dipole and quadrupole light-molecule interactions arising in surface electromagnetic fields, which exist both near rough metal and non metal surfaces. Influence of the strong quadrupole light-molecule interaction results in appearance of forbidden lines in all these processes in molecules with sufficiently high symmetry. These lines are observed in a number of molecules in all these processes. For example their existence was successfully explained in our works for ethylene, benzene, 1,3,5 trideutereobenzene, hexafluorobenzene, 1,3,5 trifluorobenzene in SERS [1], pyrazine, phenazine and 4,4'- Bipyridine in SEHRS [2,4] and in $BiPyH_2^{2+}$ and ethylene in SEIRA [3]. However as it was demonstrated in [1,5] the strong quadrupole light-molecule interaction must experience a so-called electrodynamical forbiddance in molecules with cubic symmetry groups $T, T_d, T_h, O$ and $O_h$, when all the terms, associated with the strong quadrupole light-molecule interaction become equal to zero due to belonging of the molecule to these groups and due to the electrodynamical law $div\mathbf{E} = 0$. It appears that this forbiddance must be in the fullerene molecule $C_{60}$, which belongs to the icosahedral symmetry group $Y_h$. The forbiddance must manifest in the SERS, TERS and SEIRA spectra of $C_{60}$, when the light-molecule interaction becomes of the dipole type only that results in the absence of the forbidden lines, which are usually associated with the existence of the strong quadrupole light molecule interaction.

2. **Peculiarities of the light-molecule interaction Hamiltonian in molecules with cubic and icosahedral symmetry groups.**

As it is well known, peculiarities of the spectra of optical processes in molecules are determined by the peculiarities of the light-molecule interaction Hamiltonian. For SERS, TERS and SEIRA in general one must take into account both the interaction of light with electrons and with nuclei of the molecule. In general case this Hamiltonian must be considered in the form [3]

$$\hat{H}_{mol-r} = \hat{H}_{e-r} + \hat{H}_{n-r} + \hat{H}_{nv-r} \qquad (1)$$

where

$$\hat{H}_{e-r} = |\mathbf{E}_{inc}| \frac{(\mathbf{e}^*\mathbf{f}_e^*)_{inc} e^{i\omega_{inc}t} + (\mathbf{e}\mathbf{f}_e)_{inc} e^{-i\omega_{inc}t}}{2} \qquad (2)$$

is the light-electron interaction Hamiltonian. Here $\mathbf{E}_{inc}$ is the vector of the incident electric field, $\mathbf{e}$ is the polarization vector of the incident light

$$f_{e,i} = d_{e,i} + \frac{1}{2E_i} \sum_\beta \frac{\partial E_i}{\partial x_k} Q_{e,i,k} \tag{3}$$

is the $i$ component of the generalized coefficient of the light-electron interaction,

$$d_{e,i} = \sum_j e x_{j,i} \quad , \quad Q_{e,i,k} = \sum_j e x_{j,i} x_{j,k} \tag{4}$$

are the operators of the $i$ component of the dipole moment and of the $i,k$ component of the quadrupole moments of electrons. Summation in (3) is over the index $j$, which numerates the electrons in the molecule. $\omega_{inc}$ in (2) is the frequency of the incident field. It is necessary to note that the values of the fields and their derivatives are taken in the gravity center of the molecule. In addition one should note that (3, 4) are written with the negative sign of electrons taken into account. The Hamiltonian $\hat{H}_{n-r}$ of light-nuclei interaction has the form

$$\hat{H}_{n-r} = -|\mathbf{E}_{inc}| \frac{(\mathbf{e}^*\mathbf{f}_n^*)_{inc} e^{i\omega_{inc}t} + (\mathbf{e}\mathbf{f}_n)_{inc} e^{-i\omega_{inc}t}}{2} \tag{5}$$

where

$$f_{n,i} = d_{n,i} + \frac{1}{2E_i} \sum_\beta \frac{\partial E_i}{\partial X_k} Q_{n,i,k} \tag{6}$$

is the $i$ component of generalized coefficient of the light-nuclei interaction, $d_{n,i}$ and $Q_{n,i,k}$ are the components of the dipole moment vector and of the quadrupole moments tensor of nuclei in equilibrium position, $X_k$ - is the $k$ component of the radius vectors of the gravity center of the nuclei. Since $\hat{H}_{n-r}$ is a constant, which does not depend on the electron and vibrational coordinates, then it appears that it does not define the SERS, TERS and SEIRA cross-sections and we shall not consider it further. Therefore we do not write out the values of $d_{n,i}$ and $Q_{n,i,k}$ in their evident form here.

The Hamiltonian $\hat{H}_{nv-r}$ in (1) is the one of interaction of light with the nuclei vibrations. Its form is

$$\hat{H}_{nv-r} = -\frac{|\mathbf{E}_{inc}|}{2} \times \sum_{(s,p)} \sqrt{\frac{\omega_{(s,p)}}{\hbar}} \xi_{(s,p)} [(\mathbf{e}^* \Delta \mathbf{f}_{n,(s,p)}^*)_{inc} e^{i\omega_{inc}t} + \mathbf{e}\Delta\mathbf{f}_{n,(s,p)})_{inc} e^{-i\omega_{inc}t}] \tag{7}$$

Here $\omega_{(s,p)}$ and $\xi_{(s,p)}$ are the frequency and a normal coordinate of the $(s,p)$ degenerate vibrational mode, the $s$ index numerates the groups of the degenerate states, while $p$ numerates the states inside the group.

$$\Delta f_{n,(s,p),i} = \Delta d_{n,(s,p),i} + \frac{1}{2E_{inc,i}} \sum_k \frac{\partial E_i}{\partial X_k} \Delta Q_{n,(s,p),i,k} \tag{8}$$

where

$$\Delta d_{n,(s,p),i} = \sum_J e F_J M_J \sqrt{\frac{\hbar}{\omega_{(s,p)}}} \left( \mathbf{X}_{J,(s,p)} \mathbf{e}_i \right) \quad , \tag{9}$$

$$\Delta Q_{n,(s,p),i,k} = \sum_J eF_J M_J \sqrt{\frac{\hbar}{\omega_{(s,p)}}} \left( \left(\mathbf{R}_J^0 \mathbf{e}_i\right)\left(\mathbf{X}_{J,(s,p)} \mathbf{e}_k\right) + \left(\mathbf{R}_J^0 \mathbf{e}_k\right)\left(\mathbf{X}_{J,(s,p)} \mathbf{e}_i\right) \right) \qquad (10)$$

and are interpreted as a change of the $i$ and $i,k$ components of the dipole and quadrupole moments of nuclei due to excitation of one vibrational quantum in the $(s,p)$ vibrational mode. $\mathbf{R}_J^0$ is the radius vector of the $J$ nucleus, $\mathbf{e}_k$ and $\mathbf{e}_i$ are the unit vectors in the direction of the coordinate axes, $\mathbf{X}_{J,(s,p)}$ is the displacement vector of the $J$ nucleus in the $(s,p)$ mode,

$$eF_J = \frac{eZ_J^*}{M_J} \qquad (11)$$

is the relation of the $J$ nucleus charge to its mass, which is multiplied on the electron charge. It is necessary to note that the Hamiltonian $\hat{H}_{nv-r}$ must be considered in SEIRA, but not in SERS and SEHRS, where it does not influence on these processes. Therefore first we shall consider the Hamiltonian $\hat{H}_{e-r}$.

In SERS and SEHRS on metals the enhancement is associated with the enhancement of the electric field component $E_z$, which is perpendicular to the surface near the places with a large curvature, and with the strong quadrupole light-electron interaction, which arises due to strong quadrupole transitions via the moments $Q_{e,xx}, Q_{e,yy}$ and $Q_{e,zz}$ with a constant sign and a strong increase of the derivatives $\partial E_i / \partial x_i$ of the surface electromagnetic field, which exists near a rough surface necessarily [1,2]. Indeed, the electric field near the model roughness of a wedge or a tip form with a characteristic size $l_1$ can be written approximately as [1]

$$E_r \sim |\mathbf{E}_{inc}| C_0 \left(\frac{l_1}{r}\right)^\beta , \qquad (12)$$

where $C_0 \sim 1$ is a numerical coefficient, $r$ is the distance from the observation point to the top of the wedge, or the tip. This behavior demonstrates that there is a strong enhancement of the component of the electric field, which is perpendicular to the surface near the places with a large curvature for more realistic model of the roughness as a metal tip with a finite large curvature at the top. In addition this behavior can result in the strong quadrupole interaction, which can be more than the dipole ones. Estimation of the influence of the quadrupole interaction with respect to the dipole interaction [1] is expressed as

$$\frac{\overline{\langle m|Q_{eik}|n\rangle}}{\overline{\langle m|d_{ei}|n\rangle}} \frac{1}{2E_\alpha} \frac{\partial E_i}{\partial x_k} = \frac{B_{ik} a}{2} \frac{\beta}{r} . \qquad (13)$$

Here

$$\frac{\overline{\langle m|Q_{eik}|n\rangle}}{\overline{\langle m|d_{ei}|n\rangle}} = B_{ik} a \qquad (14)$$

is the relation of some mean values of matrix elements of the quadrupole and dipole transitions, $a$ is a molecule size, $B_{ik}$ are some numerical coefficients. As it was indicated in [1], apparently the coefficients $B_{ik} \sim 1$ for $i \neq k$, since $Q_{eik}$ are of a changeable sign, while the values $B_{ii} \gg 1$, since $Q_{eii}$ are of a constant sign. The detailed reasoning one can find in [1]. The last fact results in a strong increase of the $B_{ii}$ values with respect to $B_{ik}$ $i \neq k$. Therefore the quadrupole interaction with $Q_{eii}$ moments can be significantly stronger than the dipole ones for the distances from the top of the tip

$$r < \beta \frac{B_{ii} a}{2} . \qquad (15)$$

This means that in more realistic models of the roughness, such as a tip with a finite curvature at the top, the quadrupole interaction with the moments $Q_{eii}$ can be more than the dipole ones. The qudrupole moments $Q_{e,xx}, Q_{e,yy}$ and $Q_{e,zz}$ are named as main quadrupole moments and their exceptional role in the enhancement is associated just with the fact that they are of a constant sign. The moments $Q_{e,xy}, Q_{e,xz}$ and $Q_{e,yz}$ are of a changeable sign and are non essential for the enhancement. These moments are named as minor quadrupole moments. In symmetrical molecules it is convenient to transfer to linear combinations of the quadrupole moments $Q_{e,1}, Q_{e,2}$ and $Q_{e,3}$, which transform after irreducible representations of the symmetry group of the molecule. There will be combinations with a constant sign, which are the main quadrupole moments and are essential for the enhancement and combinations with a changeable sign, which are non essential for the enhancement. After analysis of irreducible representations of all point groups it appears that the main quadrupole moments transform after the unit irreducible representation in all point groups, while the minor quadrupole moments usually transform after other irreducible representations. However there is a peculiarity of the light-electron interaction Hamiltonian in molecules with cubic and icosahedral symmetry groups $T, T_d, T_h, O, O_h$ and $Y_h$. The expressions for $Q_{e,1}, Q_{e,2}$ and $Q_{e,3}$ in them have the form

$$Q_{e,1} = \frac{1}{3}(Q_{e,xx} + Q_{e,yy} + Q_{e,zz}) \tag{16}$$

$$Q_{e,2} = \frac{1}{2}(Q_{e,xx} - Q_{e,yy}) \tag{17}$$

$$Q_{e,3} = \frac{1}{4}(Q_{e,xx} + Q_{e,yy} - 2Q_{e,zz}) \tag{18}$$

and

$$Q_{e,xx} = Q_{e,1} + \frac{2}{3}Q_{e,3} + Q_{e,2} \tag{19}$$

$$Q_{e,yy} = Q_{e,1} + \frac{2}{3}Q_{e,3} - Q_{e,2} \tag{20}$$

$$Q_{e,zz} = Q_{e,1} + \frac{2}{3}Q_{e,3} - 2Q_{e,2} \tag{21}$$

Then the value $|\mathbf{E}|(\mathbf{ef_e})$, which is contained in the light-electron interaction Hamiltonian (2) can be written as

$$|\mathbf{E}|(\mathbf{ef}_e) = (\mathbf{E}\mathbf{d_e}) + \frac{1}{2} div \mathbf{E} \times \left(Q_{e,1} + \frac{2}{3}Q_{e,3}\right) + \frac{1}{2}\left(\frac{\partial E_x}{\partial x} - \frac{\partial E_y}{\partial y} - 2\frac{\partial E_z}{\partial z}\right)Q_{e,2} + \frac{1}{2}\sum_{\substack{i,k \\ i \neq k}} \frac{\partial E_i}{\partial x_k} Q_{e,i,k} \tag{22}$$

Here $\mathbf{d_e}$ is the vector of the dipole moments of the molecule. One can see that there is only one main quadrupole moment $Q_{e,1}$, which is excluded from the Hamiltonian due to the factor $div\mathbf{E} = 0$ and does not affect on the scattering process. It is the electrodynamical forbiddance of the strong quadrupole light-electron interaction, which arises due to belonging of the molecule to the cubic, or icosahedral symmetry groups and due to the peculiarity of the strong quadrupole light-electron interaction, which becomes equal to zero due to the electrodynamical law $div\mathbf{E} = 0$.

### 3. Analysis of peculiarities of the SERS and SEIRA spectra of $C_{60}$

The scattering cross-section for the degenerate vibrational modes, which is characterized by the index $s$ in SERS in symmetrical molecules, not belonging to the indicated symmetry groups is expressed via several contributions, which describe the scattering process via various combinations of the dipole and quadrupole moments (see [1], for example).

$$d\sigma_{s,surf} = \frac{\omega_{inc}\omega_{scat}^3}{16\hbar^2\varepsilon_0^2\pi^2 c^4} \frac{|\mathbf{E}_{inc}|^2_{surf}}{|\mathbf{E}_{inc}|^2_{vol}} \frac{|\mathbf{E}_{scat}|^2_{surf}}{|\mathbf{E}_{scat}|^2_{vol}} \times$$

$$\times \sum_p \binom{(V_{(s,p)}+1)/2}{V_{(s,p)}/2} |T_{d-d} + T_{d-Q} + T_{Q-d} + T_{Q-Q}|^2_{surf} dO \quad (23)$$

Here $\omega_{inc}$ and $\omega_{scat}$ are the frequencies of the incident and scattered fields, $(\mathbf{E}_{inc})_{vol}$ and $(\mathbf{E}_{inc})_{surf}$ are the incident field in a free space and the surface field, caused by $(\mathbf{E}_{inc})_{vol}$, which affect the molecule. $(\mathbf{E}_{scat})_{vol}$ and $(\mathbf{E}_{scat})_{surf}$ are the field in a free space, which illuminates the surface from the observation point and the surface field, caused by the field $(\mathbf{E}_{scat})_{vol}$. $V_{(s,p)}$ is a vibrational quantum number of the $(s,p)$ degenerate vibrational mode, $T$ means the sum of the contributions, which arise due to the scattering via various combinations of the dipole and quadrupole moments. Here we do not write out their explicit expressions, which have a very complicated form. One can find these expressions in [1]. The sense of these contributions is that they describe the scattering process via various combinations of the dipole and quadrupole moments as it is present on Figure 1. Further we shall designate them simply as $(f_i - f_k)$, where $f_i$ and $f_k$ designate various dipole and quadrupole moments, which are involved in the scattering process. In accordance with our ideas about the role of various moments in SERS one can classify the scattering contributions after enhancement degree. It is necessary to note that the dipole interaction is essential in the enhancement also. Its enhancement arises due to the component of the electric field $E_z$, which is perpendicular to the surface. Therefore the component of the dipole moment of the molecule $d_z$, which is perpendicular to the surface is a main dipole moment. However the role of the dipole moments of the molecule in the scattering process depends on its orientation and therefore on the orientation of the dipole moment components with respect to the enhanced component of the electric field $E_z$.

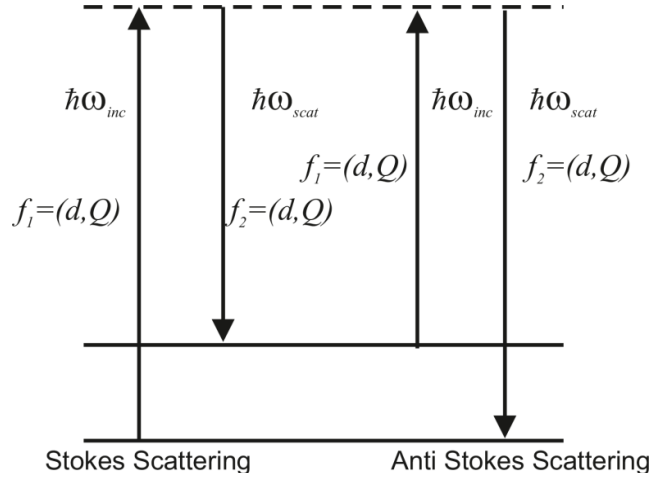

Figure 1. The scattering diagram for SERS. The scattering can arise via various combinations of the dipole and quadrupole moments $f_1$ and $f_2$.

The orientation of the molecules can be arbitrary and therefore the enhanced component of the electric field can have projections on all coordinate axes. Therefore in principle all the components of the dipole moment of the molecule can be involved in the enhancement process and can be the main ones. Therefore in general case, for symmetrical molecules, not belonging to $T, T_d, T_h, O, O_h$ and $Y_h$ symmetry groups the sequence below reflects the decrease in the enhancement degree of the scattering contributions for strong SERS:

1. the strongest, essential $(Q_{main} - Q_{main})$ scattering type;
2. essential $(d_{main} - Q_{main})$ and $(Q_{main} - d_{main})$ scattering types;
3. essential $(d_{main} - d_{main})$ scattering type.

Other contributions, which contain $Q_{\min\ or}$ quadrupole moments are non essential for the scattering and we can not to consider them here. The scattering contributions obey selection rules [1]

$$\Gamma_{(s,p)} \in \Gamma_{f_1}\Gamma_{f_2} \;\;, \qquad (24)$$

where the sign $\Gamma$ designates irreducible representations, which describe transformational properties of the vibrational mode $(s,p)$ and of the $f_1$ and $f_2$ dipole and quadrupole moments. Then, since the $Q_{main}$ moments transform after the unit irreducible representation, in accordance with the selection rules (24), the most enhanced contributions determine the most enhanced lines, which are associated with the vibrations, transforming after the unit irreducible representation. The enhanced contributions of the $(d_{main} - d_{main})$ scattering type with the same indices of the dipole moment components, designated here as *main*, contribute to these lines also. However the contributions of the $(d_{main} - Q_{main})$ and $(Q_{main} - d_{main})$ scattering types determine appearance of the lines, caused by vibrations, transforming as the $d_{main}$ moments. These lines may be forbidden in usual Raman scattering and IR active in usual IR absorption in molecules where the $d_{main}$ moments transform after the irreducible representations, which are not the unit one.

Further we shall consider only the regularities of the experimental SERS spectra of the fullerene $C_{60}$ since consideration of other cubic molecules requires a specific consideration of the symmetry group properties and the results for the cubic molecules and the fullerene $C_{60}$ can differ one from another.

In molecules with the cubic symmetry groups and in the fullerene $C_{60}$ the first and the second types of the contributions are forbidden since the $Q_{main}$ quadrupole moments are excluded due to the electrodynamical forbiddance. Therefore this fact must result in the absence of the forbidden lines, associated with the $(d_{main} - Q_{main})$ and $(Q_{main} - d_{main})$ scattering contributions specifically for $C_{60}$. The SERS spectrum is defined only by the dipole interaction such as in a usual Raman scattering. Therefore the maximum enhancement in SERS in molecules with cubic symmetry groups and in the fullerene $C_{60}$ can be lower than in SERS in other molecules, not belonging to the cubic and icosahedral symmetry groups since it is associated with the enhancement of the electric field only. This result was observed and indicated in SERS on methane [5] and in Single molecule SERS on $C_{60}$ [6] where the estimated enhancement coefficient was $\sim 10^8$ instead of $10^{14} - 10^{15}$ in Single molecule SERS on other molecules [7].

It is necessary to note that SERS at present is used in a special spectroscopic technic named as TERS (Tip enhanced spectroscopy). It is SERS in fact, which uses a metal tip as a surface roughness. Therefore main regularities of the TERS spectra are the same as in SERS and there must be forbidden lines, caused by $(Q_{main} - d_{main})$ scattering type, which can be inactive in usual Raman scattering and are active in a usual IR absorption. Because of the electrodynamical forbiddance, these lines must be absent in SERS and TERS spectra of $C_{60}$. Analysis of the SERS and TERS spectra of this molecule obtained in [8] indicates that there are only the lines, caused by the scattering of the $(d_i - d_k)$ type. As an example we present one of the SERS spectra, obtained in [8] (Figure 2).

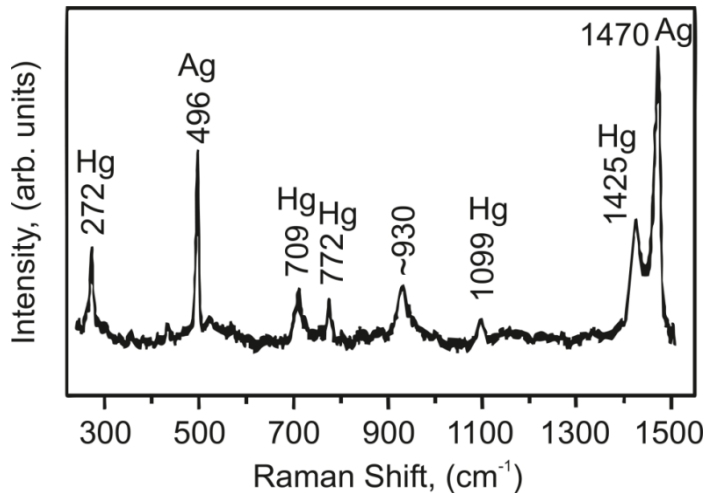

Figure 2. The SERS spectrum of $C_{60}$. The lines, caused by the vibrations with the $T_{1u}$ irreducible representation at 526, 575, 1182 and 1429 $cm^{-1}$, which are allowed in usual IR and in SERS in other symmetrical molecules are forbidden in $C_{60}$ due to the electrodynamical forbiddance of the strong quadrupole light-molecule interaction.

One can see the lines at $496 cm^{-1}$ and $1470 cm^{-1}$ caused by totally symmetric vibrations with the unit irreducible representation $A_g$ and by the $(d_i - d_i)$ scattering type and the lines at 273, 435, 710, 774, 1098, and 1425 $cm^{-1}$, caused by the vibrations with the irreducible representation $H_g$, associated with the $(d_i - d_k)$ $(i \neq k)$ types of the scattering. The lines at 526, 575, 1182 and 1429 $cm^{-1}$, which are associated with the vibrations with the $T_{1u}$ irreducible representation describing transformational properties of the dipole moments are forbidden and are not observed in all the spectra, presented in [8]. It is necessary to note existence of the line at $930 cm^{-1}$ on Figure 2. This line does not belong to the lines of $C_{60}$ and is absent in other SERS and TERS spectra, published in [8]. Therefore its appearance may be associated with contamination of the sample. Thus the experimental SERS and TERS spectra of the fullerene $C_{60}$ are in a full agreement with our ideas about existence of the electrodynamical forbiddance of the strong quadrupole light-molecule interaction.

It is necessary to note that the electrodynamical forbiddance in cubic molecules like methane in SERS was predicted in [9,10] and then in SEIRA in [11]. It must be observed in the SEIRA spectra of $C_{60}$ also. As it has been pointed out above the forbidden lines in SEIRA, caused by vibrations with the unit irreducible representation were predicted in [12] and were observed in ethylene and the ion $BiPyH_2^{2+}$ with the $D_{2h}$ symmetry group [13,14]. However such lines must be forbidden in $C_{60}$. Here we do not analyze the influence of $\hat{H}_{nv-r}$ Hamiltonian (7), which must be analyzed for explanation of appearance of forbidden lines in SEIRA. This analysis was made partially in [11] and it was demonstrated that the electrodynamical forbiddance is valid in SEIRA when the Hamiltonian $\hat{H}_{nv-r}$ is taken into account too. It must manifest in the absence of the lines, associated with the totally symmetric vibrations, transforming after the unit irreducible representation, which can appear in molecules with other symmetry groups that the cubic and icosahedral ones. Here we shall consider only manifestation of the electrodynamical forbiddance in the experimental SEIRA spectra of $C_{60}$ adsorbed on a gold substrate, which were obtained in [15]. The authors investigated the spectra of two specimens with fullerenes $C_{60}$, obtained by various technological methods. One of the samples was named as FWS (fullerene-water-system) and was characterized as an aqueous molecular-colloidal solution, which contained both single fullerene molecules and their fractal clusters in hydrated state. The second was a monodisperse aqueous colloidal system, which is a typical $C_{60}$ hydrosol with the size of the particles $\sim 10 nm$. It was named as ChH. In [15] it was demonstrated that for FWS the line at $1470 cm^{-1}$, caused by the totally symmetric vibration with the unit irreducible representation $A_g$ is absent (Figure 3). Regretfully the SEIRA spectrum was obtained in the interval, which is above the value of the wavenumber $500 cm^{-1}$. Therefore the second line with the unit irreducible representation at $496 cm^{-1}$ is not within the interval measured. However in principle it could be seen at the wavenumber $500 cm^{-1}$ because of a possible sufficiently large line width. However no such data were obtained in [15]. In addition all the lines at 526, 275, 1182 and 1429 $cm^{-1}$ characteristic for usual IR absorption and SEIRA and associated with the vibrations with the $T_{1u}$ irreducible representation describing transformational properties of the dipole moments of $C_{60}$ are present. It is necessary to note existence of some lines at $615 cm^{-1}$ (the symmetry is not defined), $669 cm^{-1}$ and $747 cm^{-1}$ with the $H_u$ symmetry and at $833 cm^{-1}$ with the $T_{1g}$ symmetry, which either not belong to $C_{60}$, or are forbidden in its SEIRA spectrum. Appearance of these lines, apparently is associated with not very pure experimental conditions, when the fullerene molecules can be distorted by an environment or there may be by other reasons like contamination. However the absence of the lines with the $A_g$ irreducible representation and existence of the lines, caused by the pure dipole interaction confirm correctness of our results and the existence of the electrodynamical forbiddance. One should note that the SEIRA spectrum of the ChH system contains the broad line at $1462 - 1465 cm^{-1}$, which is very close to the line of $C_{60}$ at $1470 cm^{-1}$ with the $A_g$ irreducible representation that contradicts to our ideas. However, from our opinion the existence of this line apparently is associated with "not very pure system" and may be due to possible presence of the fullerenes $C_{70}$ where there is the line with the wavenumber, which is very close to $1470 cm^{-1}$.

Thus we have considered the SERS, TERS and SEIRA experiments on the fullerene $C_{60}$, which confirm experimentally with a large probability existence of the electrodynamical forbiddance of the strong quadrupole light-molecule interaction.

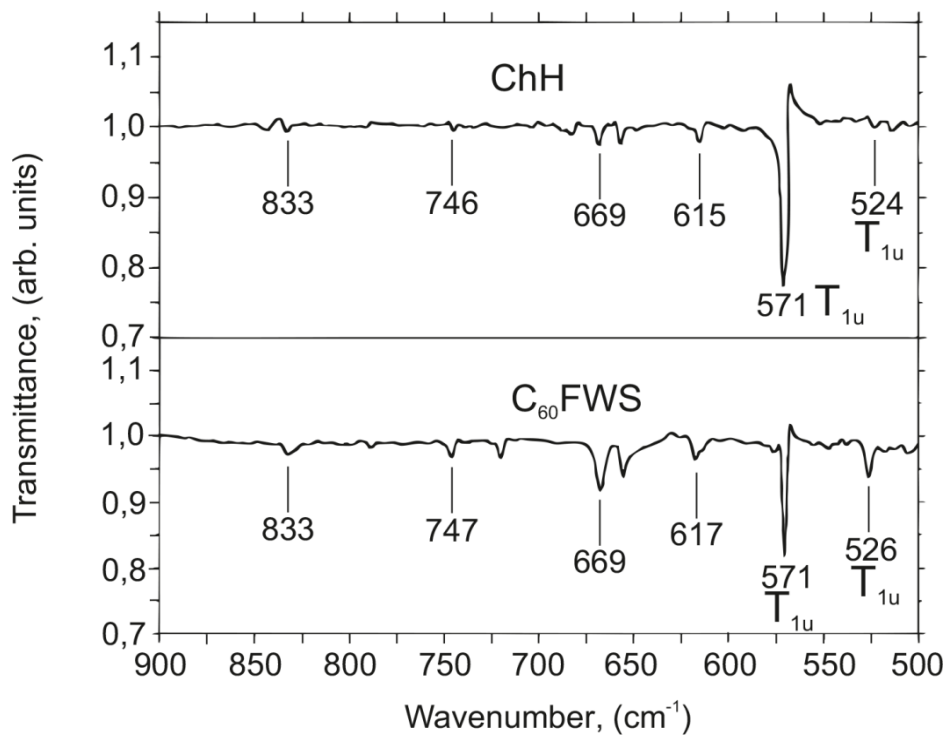

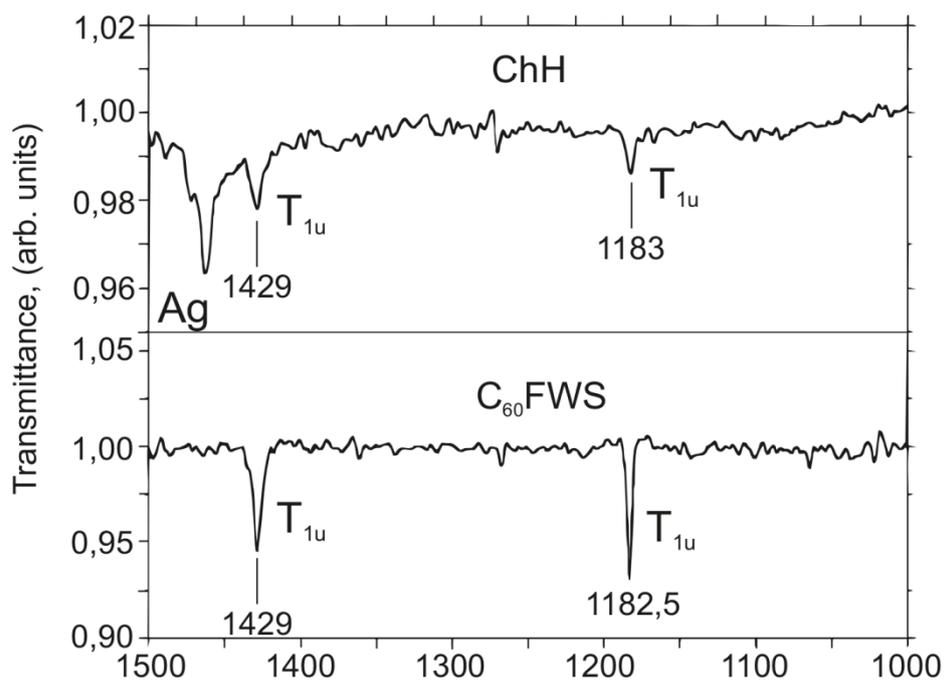

Figure 3. The SEIRA spectra of the fullerene $C_{60}$ for two types of the samples, FWS and ChH. The SEIRA spectrum for FWS demonstrates the absence of the lines, caused by the totally symmetric vibrations with the unit irreducible representation $A_g$ at $1470 cm^{-1}$ that indicated on the electrodynamical forbiddance of the strong quadrupole light-molecule interaction in SEIRA.